\documentclass[10pt,conference]{IEEEtran}
\usepackage{graphicx}
\usepackage[dvips]{color}
\usepackage{epsfig}
\usepackage{setspace}
\usepackage{amssymb}
\usepackage{amsmath}

\begin{document}
\title{On General Lattice Quantization Noise}

\author{
\authorblockN{Tal Gariby and  Uri Erez
\\[1mm]
         Department of Elect. Eng. - Systems, Tel-Aviv
        University
        }
}
\maketitle

\begin{abstract}
The problem of constructing lattices such that their quantization noise approaches a desired
distribution is studied. It is shown that asymptotically in the dimension, lattice quantization
noise can approach a broad family of distribution functions with independent and identically
distributed components.
\end{abstract}


\def\N{{\rm I\mkern-3mu N}}
\def\R{{\rm I\mkern-3mu R}}
\def\Q{{\@QC Q}}
\def\C{{\@QC C}}
\def\@QC#1{\mathpalette{\setbox0=\hbox\bgroup$\rm}%
  {\egroup C$\egroup\rm\rlap{\kern0.4\wd0\vrule
  width 0.05\wd0 height 0.97\ht0 depth -0.0
1\ht0}%
  #1\bgroup}}

\newcommand{\modulo}{\mbox{\,mod\,}}
\newcommand{\neff}{N_{\rm eff}}
\newcommand{\ebj}{e^{j 2 \pi f}}
\newcommand{\uzer}{\underline{0}}
\newcommand{\uV}{\underline{V}}
\newcommand{\uA}{\underline{A}}
\newcommand{\uD}{\underline{D}}
\newcommand{\uv}{\underline{v}}
\newcommand{\uT}{\underline{T}}
\newcommand{\ut}{\underline{t}}
\newcommand{\ur}{\underline{r}}
\newcommand{\uR}{\underline{R}}
\newcommand{\uc}{\underline{c}}
\newcommand{\uC}{\underline{C}}
\newcommand{\ul}{\underline{l}}
\newcommand{\uL}{\underline{L}}
\newcommand{\uh}{\underline{h}}
\newcommand{\uH}{\underline{H}}
\newcommand{\ue}{\underline{e}}
\newcommand{\uE}{\underline{E}}
\newcommand{\uG}{\underline{G}}
\newcommand{\ug}{\underline{g}}
\newcommand{\uz}{\underline{z}}
\newcommand{\uZ}{\underline{Z}}
\newcommand{\uu}{\underline{u}}
\newcommand{\uU}{\underline{U}}
\newcommand{\uj}{\underline{j}}
\newcommand{\uJ}{\underline{J}}
\newcommand{\uX}{\underline{X}}
\newcommand{\ux}{\underline{x}}
\newcommand{\uY}{\underline{Y}}
\newcommand{\uy}{\underline{y}}
\newcommand{\uW}{\underline{W}}
\newcommand{\uw}{\underline{w}}
\newcommand{\uth}{\underline{\theta}}
\newcommand{\uTh}{\underline{\Theta}}
\newcommand{\uph}{\underline{\phi}}
\newcommand{\ual}{\underline{\alpha}}
\newcommand{\uxi}{\underline{\xi}}
\newcommand{\us}{\underline{s}}
\newcommand{\uS}{\underline{S}}
\newcommand{\un}{\underline{n}}
\newcommand{\uN}{\underline{N}}
\newcommand{\up}{\underline{p}}
\newcommand{\uq}{\underline{q}}
\newcommand{\uf}{\underline{f}}
\newcommand{\ua}{\underline{a}}
\newcommand{\ub}{\underline{b}}
\newcommand{\uDelta}{\underline{\Delta}}
\newcommand{\cA}{{\cal A}}
\newcommand{\cB}{{\cal B}}
\newcommand{\cC}{{\cal C}}
\newcommand{\cc}{{\cal c}}
\newcommand{\cD}{{\cal D}}
\newcommand{\cE}{{\cal E}}
\newcommand{\cF}{{\cal F}}
\newcommand{\cI}{{\cal I}}
\newcommand{\cJ}{{\cal J}}
\newcommand{\cK}{{\cal K}}
\newcommand{\cL}{{\cal L}}
\newcommand{\cN}{{\cal N}}
\newcommand{\cP}{{\cal P}}
\newcommand{\cQ}{{\cal Q}}
\newcommand{\cR}{{\cal R}}
\newcommand{\cS}{{\cal S}}
\newcommand{\cs}{{\cal s}}
\newcommand{\cT}{{\cal T}}
\newcommand{\ct}{{\cal t}}
\newcommand{\cU}{{\cal U}}
\newcommand{\cV}{{\cal V}}
\newcommand{\cW}{{\cal W}}
\newcommand{\cX}{{\cal X}}
\newcommand{\cx}{{\cal x}}
\newcommand{\cY}{{\cal Y}}
\newcommand{\cy}{{\cal y}}
\newcommand{\cZ}{{\cal Z}}
\newcommand{\tE}{\tilde{E}}
\newcommand{\tG}{\tilde{G}}
\newcommand{\tZ}{\tilde{Z}}
\newcommand{\tz}{\tilde{z}}
\newcommand{\hU}{\hat{U}}
\newcommand{\hX}{\hat{X}}
\newcommand{\hY}{\hat{Y}}
\newcommand{\hZ}{\hat{Z}}
\newcommand{\huX}{\hat{\uX}}
\newcommand{\huY}{\hat{\uY}}
\newcommand{\huZ}{\hat{\uZ}}
\newcommand{\indp}{\underline{\; \| \;}}
\newcommand{\diag}{\mbox{diag}}
\newcommand{\sumk}{\sum_{k=1}^{K}}
\newcommand{\beq}[1]{\begin{equation}\label{#1}}
\newcommand{\eeq}{\end{equation}}
\newcommand{\req}[1]{(\ref{#1})}
\newcommand{\figref}[1]{Figure \ref{#1}}
\newcommand{\secref}[1]{Section \ref{#1}}
\newcommand{\thref}[1]{Theorem \ref{#1}}
\newcommand{\beqn}[1]{\begin{eqnarray}\label{#1}}
\newcommand{\eeqn}{\end{eqnarray}}
\newcommand{\limn}{\lim_{n \rightarrow \infty}}
\newcommand{\limN}{\lim_{N \rightarrow \infty}}
\newcommand{\limr}{\lim_{r \rightarrow \infty}}
\newcommand{\limd}{\lim_{\delta \rightarrow \infty}}
\newcommand{\limM}{\lim_{M \rightarrow \infty}}
\newcommand{\limsupn}{\limsup_{n \rightarrow \infty}}
\newcommand{\imii}{\int_{-\infty}^{\infty}}
\newcommand{\imix}{\int_{-\infty}^x}
\newcommand{\ioi}{\int_o^\infty}
\newcommand{\vecN}{_0^{N-1}}

\newcommand{\bphi}{\mbox{\boldmath \begin{math}\phi\end{math}}}
\newcommand{\bth}{\mbox{\boldmath \begin{math}\theta\end{math}}}
\newcommand{\bhth}{\mbox{\boldmath \begin{math}\hat{\theta}\end{math}}}
\newcommand{\bg}{\mbox{\boldmath \begin{math}g\end{math}}}
\newcommand{\ba}{{\bf a}}
\newcommand{\bb}{{\bf b}}
\newcommand{\bl}{{\bf l}}
\newcommand{\bc}{{\bf c}}
\newcommand{\br}{{\bf r}}
\newcommand{\bD}{{\bf D}}
\newcommand{\RR}{{\mathbb{R} }}
\newcommand{\ZZ}{{\mathbb{Z} }}
\newcommand{\bbf}{{\bf f}}
\newcommand{\bn}{{\bf n}}
\newcommand{\bs}{{\bf s}}
\newcommand{\bd}{{\bf d}}
\newcommand{\bt}{{\bf t}}
\newcommand{\bu}{{\bf u}}
\newcommand{\bx}{{\bf x}}
\newcommand{\bw}{{\bf w}}
\newcommand{\by}{{\bf y}}
\newcommand{\bz}{{\bf z}}
\newcommand{\bI}{{\bf I}}
\newcommand{\bC}{{\bf C}}
\newcommand{\bE}{{\bf E}}
\newcommand{\bJ}{{\bf J}}
\newcommand{\bN}{{\bf N}}
\newcommand{\bU}{{\bf U}}
\newcommand{\bQ}{{\bf Q}}
\newcommand{\bS}{{\bf S}}
\newcommand{\bT}{{\bf T}}
\newcommand{\bV}{{\bf V}}
\newcommand{\bW}{{\bf W}}
\newcommand{\bX}{{\bf X}}
\newcommand{\bY}{{\bf Y}}
\newcommand{\bZ}{{\bf Z}}
\newcommand{\byt}{{\bf y_t}}
\newcommand{\oI}{\overline{I}}
\newcommand{\oD}{\overline{D}}
\newcommand{\oh}{\overline{h}}
\newcommand{\oV}{\overline{V}}
\newcommand{\oR}{\overline{R}}
\newcommand{\oH}{\overline{H}}
\newcommand{\ol}{\overline{l}}
\newcommand{\E}{{\cal E}_d}
\newcommand{\el}{\ell}
\newcommand{\snr}{{\rm SNR}}

\newcommand{\var}{{\rm Var}}

\newcommand{\bneff}{\bN_{\rm eff}}

\newtheorem{theorem}{Theorem}


\newcommand{\yesindent}{\hspace*{\parindent}}   
\newtheorem{thmbody}{Theorem}
\newenvironment{thm}{
    \begin{singlespace} \begin{thmbody}
    }{
    \end{thmbody} \end{singlespace}
    }
\newtheorem{dfnbody}{Definition}
\newenvironment{dfn}{
    \begin{singlespace} \begin{dfnbody}
    }{
    \end{dfnbody} \end{singlespace}
    }
\newtheorem{corbody}{Corollary}
\newenvironment{cor}{
    \begin{singlespace} \begin{corbody}
    }{
    \end{corbody} \end{singlespace}
    }
\newtheorem{lemmabody}{Lemma}
\newenvironment{lemma}{
    \begin{singlespace} \begin{lemmabody}
    }{
    \end{lemmabody} \end{singlespace}
    }
\newtheorem{propbody}{Proposition}
\newenvironment{prop}{
    \begin{singlespace} \begin{propbody}
    }{
    \end{propbody} \end{singlespace}
    }
\newenvironment{example}{
    \begin{small} \begin{singlespace}{\it Example:}
    }{
    \end{singlespace} \end{small}
    }
\newcommand{\pderiv}[2]{\frac{ \partial {#1}}{ \partial {#2}}}
\newcommand{\overr}[2]{\left({\begin{array}{l}#1\\#2\end{array}}\right)}
 \newcommand{\Ddef}{\stackrel{\Delta}{=}}



\section{Introduction}
Lattices play a key role in digital communication and specifically
in quantization theory. In high-resolution quantization theory, it
is common to assume (see, e.g, \cite{NeuhoffGray98} and references therein)
that the quantization error of a lattice
quantizer is uniformly distributed over the basic cell of the
lattice. This assumption can be made completely accurate at any
resolution by means of subtractive dithered quantization, where a random vector
(dither) which is uniformly distributed over the lattice cell is
added prior to quantization and then subtracted from the quantizer
output (see \cite{DitheredQuant}). Following \cite{FederZamirLQN} we
thus use the term {\em lattice quantization noise} (LQN)  to refer
to a random vector uniformly distributed over a basic cell of the
lattice. For a given lattice however there are many possible
partitions into cells. Each such partition will result in a
different LQN with different statistical properties (see
\cite{AsymDist}).

In many cases of interest, the criterion for quantization is that of
minimum mean-square error (MSE). That is, a lattice is deemed
good if the MSE of the quantization noise (resulting from nearest neighbor encoding) is minimal for a given
lattice density (or cell volume). When the
criterion is MSE, the basic region associated with nearest neighbor encoding (in the Euclidian sense) is called the Voronoi
region. The statistical properties of LQN
of lattices which are good in this sense has been thoroughly
investigated by Zamir and Feder \cite{FederZamirLQN}. Indeed, it was shown in \cite{FederZamirLQN} that there exist sequences of
lattices which are asymptotically optimal in an MSE sense. That is, for such sequences, the
normalized second moment of the Voronoi region of the lattice goes
to $\frac{1}{2 \pi e}$ as the dimension goes to infinity and the distribution of quantization noise (over
a Voronoi region) approaches (in the Kullback-Leibler divergence
sense) that of an i.i.d. white Gaussian noise.

In certain cases, the criterion for quantization may be different from MSE. For instance
one may be interested in some other metric, e.g., an $r$-th power norm. More generally, one may ask
whether one can construct a lattice and associate with it a {\em lattice partition} such that the
corresponding LQN approaches {\em any} i.i.d. distribution. The interest of the authors in this question
arose when general LQN was needed in the context of designing a lattice precoding scheme for the binary
dirty paper problem \cite{GES}.

The results of \cite{FederZamirLQN} were derived using previously known results
\cite{Rogers} on the existence of lattices that are good for the classical problem of
covering. This approach unfortunately does not lend itself to
extending the results to more general distributions. On the other
hand,  typicality arguments and rate distortion theory suggest that
a random code drawn uniformly over a large region
should have the desired properties. Since linear codes and
lattices have proved to be able to attain the performance of a
(uniform) random code in many problems in information theory, it is
natural to suspect that the same would hold for the problem at hand.

Indeed, one may use random coding (or averaging arguments) to obtain existence results for lattices, an approach
dating back at least as far as Hlawka's proof of the Minkowsk-Hlawka theorem, see \cite{Loeliger} for an historical
account and further details.
In \cite{Loeliger}, Loeliger defined an ensemble of lattices based on Construction~A (see \cite{ConwaySloane}) which is
very amenable to analysis. Loeliger then used averaging and typicality arguments to obtain channel coding theorems for lattice codes.

In \cite{ELZ} the Loeliger ensemble (with a careful choice of parameters)
was used to establish the existence
of lattices that are simultaneously good under various different notions. It was further noted in
\cite{ELZ} that the results can be extended to show that there exist lattices that are good for
quantization under any $r$-th norm.
In this work we extend these results to show that under quite
general conditions, LQN can approach general i.i.d. distributions. In the proof we use the same ensemble of lattices as in \cite{Loeliger} and \cite{ELZ}. However, the proof technique diverges
from that of \cite{ELZ} in that it relies on typicality as in  \cite{Loeliger} rather than geometric arguments to form the
lattice cells. In this sense the present work is dual to Loeliger's work \cite{Loeliger} , using typicality arguments
to obtain results for source coding (rather than for channel coding as in \cite{Loeliger}).

The paper is organized as follows.
Section \ref{sec:Pre} provides a very brief introduction  to lattices and lattice quantization noise as well as
states the main result of the paper for both the discrete and continuous cases.
Section \ref{sec:ensemble} describes the ensemble of lattices to be used and defines the typicality-based lattice partition.
Sections~\ref{sec:proof} and \ref{sec:tying} provides the proof for the existence of a lattices whose quantization noise approaches a desired
distribution.  Finally, the results are demonstrated in Section \ref{sec:sim} by simulation, finding
lattices and partitions with quite arbitrary LQN noise.

\section{Preliminaries and statement of main result}
\label{sec:Pre}
\subsection{Lattices and Lattice Quantization Noise}
We begin by recalling a few basic notions pertaining to lattices. An $n$-dimensional lattice
$\Lambda$ is an infinite discrete subgroup of the Euclidean space $\mathbb{R}^n$. Thus, if
$\lambda_1$ and $\lambda_2$ are in $\Lambda$, then their sum and difference are also in $\Lambda$.
An $n$-dimensional lattice may be defined by an $n \times n$ generating matrix $\tG$ (whose choice is not unique) such that

\[\Lambda = \{\by:  \by= \bx \cdot \tG  {\rm \ for \  some\ }  \bx \in \mathbb{Z}^n\} .\]

We may associate with a lattice $\Lambda$ a {\em lattice partition}, partitioning $\RR^n$ into
disjoint cells. We denote by $\cV=\cV_0$ the fundamental cell associated with $\lambda=0$. We
further associate with every lattice point $\lambda$ the cell $\cV_\lambda \Ddef \lambda+\cV$.
There are many possible choices for $\cV$. For $\cV$ to be valid however, we require that every
point $\by \in \mathbb{R}^n$ can be {\em uniquely} written as $\by = \lambda + \br $ where
$\lambda \in \Lambda$ is a lattice point and $\br \in \cV$ is the ``remainder". We may thus write
$\mathbb{R}^n = \Lambda + \cV$. The lattice partition is therefore fully determined by the
specification of the fundamental region $\cV$. The volume of a fundamental region is the same for any
valid partition and we denote it by $V(\Lambda)$ or simply by $V$.

\begin{figure}[htb]
\epsfysize=1.1in
\includegraphics[width=0.9\columnwidth]{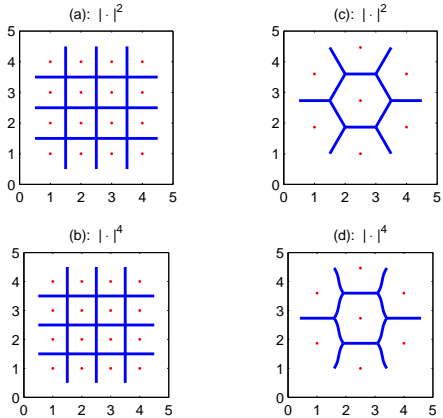}
\caption{Examples of lattices and lattice partitions: (a) The lattice is $\ZZ^n$ with Voronoi partitioning corresponding to the Euclidean norm.
(b) The lattice is $\ZZ^n$ with partitioning corresponding to the $4$-power norm, which still results in Voronoi partitioning.
(c) Hexagonal lattice with Voronoi partitioning. (d) Hexagonal lattice with partitioning corresponding to the $4$-power norm.}
\label{partitions}
\end{figure}

We note that when the partition is such that a
point is mapped to the nearest lattice point in $\Lambda$ in the Euclidean sense, we obtain the
usual Voronoi partition. An example of lattices and lattice partitions is given in Figure~\ref{partitions}.


We further associate with a lattice and a chosen partition $\{\Lambda,\cV \}$ a lattice quantizer
$Q_\cV(\cdot)$. For any $\by \in \mathbb{R}^n$, since it may be written as $\by=\lambda+\br$
in one and only way, we define $Q_\cV(\by)=\lambda$ to be the quantization of $\by$. We further
define a modulo operation by,
\begin{eqnarray*}
\by \modulo \Lambda  \Ddef \by - Q_\cV(\by)  =  \br.
\end{eqnarray*}
For any input vector $\by$, we may view the remainder $\br=\by \modulo \Lambda$ as the ``quantization noise" associated with $\by$. A random vector ${\bf U}$ uniformly distributed over $\cV$ is referred to as (random) LQN.

\subsection{Statement of main result}
Let $\bW$ be an i.i.d. random vector with marginal Probability Density Function (PDF) denoted by $f_W(w)$. Then we
would like to find a sequence of lattices and corresponding partitions such that the associated LQN approaches an i.i.d.
distribution with marginal PDF $f_W(w)$.

\begin{dfn}
\label{def1_cont}
A random variable $W$ with a continuous PDF $f_W(\cdot)$ will be called permissible if $f_W(\cdot)$ is bounded from below by a positive number over a closed interval $\cA=[-A,A]$, and is zero outside of $\cA$, i.e,
\begin{eqnarray}
f_W(w): \cA \longrightarrow [a_{\rm min},a_{\rm max}] \label{FuncCond}
\end{eqnarray}
where $a_{\rm min}>0$ is some arbitrarily small fixed value, and $a_{\rm max}$ is some arbitrarily
large fixed value.
\end{dfn}

\begin{thm}
\label{ContinuousShaping} Let $W$ be a permissible noise with PDF $f_W(\cdot)$. Let $\bW$ be drawn i.i.d. $\sim \prod f_W$. Then, there exists a sequence of lattices and associated partitions such that the resulting lattice quantization noise $\bU$ satisfies,
\begin{eqnarray}
\limsup_{n \rightarrow \infty} \frac{1}{n}D(\bU||\bW) = \xi
\end{eqnarray}
where $D(\cdot \| \cdot)$ is the Kullback-Leibler divergence and $\xi$ can be taken
arbitrarily small.
\end{thm}

This theorem also results in the following corollary that shows convergence of the {\em marginal}
PDFs to the desired one (in an average sense).
\begin{cor}
\label{OneDimensionCont} Let $W$ be a permissible noise with PDF $f_W(\cdot)$.
Let $\bW$ be drawn i.i.d. $\sim \prod f_W$. Then there exists a sequence of lattices
and associated partitions such that the resulting
lattice quantization noise, $\bU$ satisfies,
\begin{eqnarray}
\limsup_{n \rightarrow \infty} \frac{1}{n} \sum_{i=1}^{n} D(U_i||W) = \xi
\end{eqnarray}
where $\xi$ can be taken arbitrarily small.
\end{cor}

The key to proving this theorem lies in proving a similar claim for the discrete case which we
state next. The proof for the continuous case follows from the discrete case by standard
arguments, dividing the interval $\cA$ into small enough intervals and is given in Section~\ref{sec:tying}.

\subsection{Discrete case}
\label{sec:disc}
We now restrict our attention to the discrete space
$\mathbb{Z}^n$. We begin by recalling some basic definitions which are analogous to
those provided above for the continuous setting.

A fundamental cell $\cV$ associated with the lattice
$\Lambda \subset \mathbb{Z}^n$ is a finite set $\cV \subset
\mathbb{Z}^n$ such that any point $\by \in \mathbb{Z}^n$ can be
written in one and only one way as $\by =\br+\lambda$ where $\lambda
\in \Lambda$ and $\br \in \cV$. We further define the corresponding
quantizer and LQN as before. 

Let $p$ be a prime number. The following notation will be used in the paper to denote componentwise modulo operations:
\begin{itemize}
\item For any scalar random variable $X$, $X^* = X \modulo p$.
\item For any vector random variable $\bX$, $\bX^*$ is the result of reducing each component of $\bX$ modulo $p$.
\end{itemize}
The discrete counterparts of
Definition~\ref{def1_cont}, Theorem~\ref{ContinuousShaping}, and Corollary~\ref{OneDimensionCont} are:

\begin{dfn}
\label{def1_discrete}
A random variable $W$ with a discrete probability function $P_W(\cdot)$ will be called $p$-permissible if $W$ takes values in $\mathbb{Z}_p = \{0,...,p-1\}$ and $P_W(\cdot)$ is a strictly positive probability function, i.e.,
\begin{eqnarray*}
P_W(w):  \mathbb{Z}_p \longrightarrow [a_{\rm min},1),
\end{eqnarray*}
where $a_{\rm min}$ (with abuse of notation) is some arbitrarily small fixed value.
\end{dfn}

\begin{thm}
\label{DiscreteShaping} Let $W$ be a $p$-permissible noise with a PDF $P_W(\cdot)$ where $p$ is a prime number. Let $\bW$ be drawn i.i.d. $\sim \prod P_W$. Then, there exists a sequence of integer valued lattices and associated partitions such that the resulting lattice quantization noise, $\bU$, satisfies,
\begin{eqnarray}
\lim_{n \rightarrow \infty} \frac{1}{n}D(\bU^*||\bW) = 0.
\end{eqnarray}
\end{thm}

\begin{cor}
\label{OneDimensionDisc} Let $W$ be a $p$-permissible noise with a PDF $P_W(\cdot)$ where $p$ is a prime number. Let $\bW$ be drawn i.i.d. $\sim \prod P_W$. Then there exists a sequence of
lattices and associated lattice partitions such that the resulting lattice quantization noise, $\bU$, satisfies
\begin{eqnarray}
\lim_{n \rightarrow \infty} \frac{1}{n} \sum_{i=1}^{n} D(U_i^*||W) = 0.
\end{eqnarray}
\end{cor}

The proof of Theorem \ref{DiscreteShaping} is based on defining an appropriate ensemble of
lattices as in \cite{ELZ}, and then defining quantization cells based on a typicality ``metric".
The proof of Corollary~\ref{OneDimensionDisc} is given in Appendix \ref{sec:proof_of_corollary}.

\section{Ensemble of Lattices and Lattice Partition}
\label{sec:ensemble}

\begin{figure}[htb]
\epsfysize=1.1in
\includegraphics[width=0.9\columnwidth]{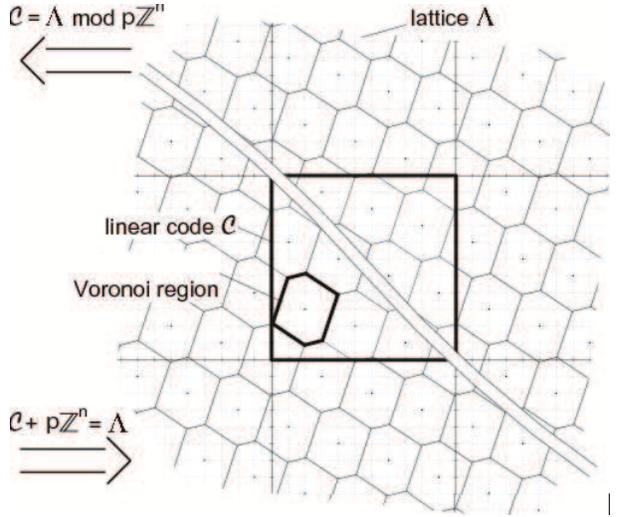}
\caption{Depiction of Construction~A: A lattice is obtained by replicating a linear code $\cC \subset \ZZ_p^n$ over
$\RR^n$.} \label{CONA}
\end{figure}

We make use of the Loeliger ensemble of lattices \cite{Loeliger} based on Construction~A (see \cite{ConwaySloane}).
Let $k$, $n$, be integers such that $k
< n$ and let $p$ be a prime number. Let $G$ be a $k \times n$ generating
matrix with elements in $\mathbb{Z}_p=\{0,1,\ldots,p-1\}$. Then a  lattice may be obtained from $G$ using Construction~A as
depicted in Figure~\ref{CONA}. The construction consists of the
following steps:
\begin{itemize}
\item Define the codebook $\cC=\{\bx=\bu \cdot G : \bu \in \mathbb{Z}_p^k\}$,
where all the operations are modulo-$p$. Thus $\cC \subset \mathbb{Z}_p^n$.
The rate of the code\footnote{All logarithms in this paper are taken to base $2$.} (in bits per sample), $R$, is defined by
\begin{eqnarray}
\label{Rdef}
R = \frac{\log p^k}{n}
\end{eqnarray}
\item Replicate $\cC$ over $\mathbb{Z}^n$ to form the lattice $\Lambda = \cC+p\mathbb{Z}^n$.
It is easy to show that $\Lambda$ is indeed a lattice, see, e.g., \cite{ConwaySloane}.
\end{itemize}
The {\em random}  ensemble of lattices is generated by drawing
each entry of the generating matrix $G$ according to a uniform
i.i.d. distribution over $\mathbb{Z}_p$, resulting in a random codebook\footnote{We note that with this notation (numbering), some of
the codewords may be identical (when $G$ is not full rank). That is of no consequence to the analysis.}
 $\cC=\{ \bX_1, \bX_2, \ldots, \bX_{p^k} \}$, and applying the steps
described above. 

\begin{figure}[htb]
\epsfysize=1.1in
\includegraphics[width=0.99\columnwidth]{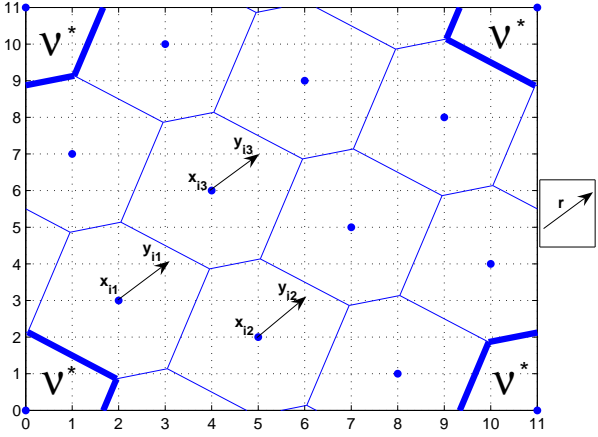}
\caption{Forming the lattice partition} \label{code_mapping}
\end{figure}

We note that Construction~A results in a lattice that is periodic with respect to the lattice $p\mathbb{Z}^n$ as can be seen in Figure~\ref{CONA}. Thus, when analyzing the properties of the lattice, one may restrict attention to the the basic cube (the region highlighted in Figure~\ref{CONA}), i.e., to the region $\mathbb{Z}_p^n$.
In particular, for any lattice $\Lambda$ and fundamental region $\cV$, we define the ``folded" fundamental region $\cV^* = \cV \modulo p$
as depicted in Figure~\ref{code_mapping}.
It is easy to see that $\cV^*$ plays the same role with respect to the code $\cC=\Lambda \modulo p$ as $\cV^*$ does
 with respect to $\Lambda$. That is, for each $\by \in \mathbb{Z}_p^n$ one may write $\by = (\bx + \br) \modulo p$ where $\bx \in \cC$ and $\br \in \cV^*$. Note also that in the same manner that specifying $\cV$ induces the folded region $\cV^*$, the converse is also true.
Specifying the region $\cV^*$ with respect to the code $\cC$, naturally induces the region $\cV$ with respect to the lattice.
Further, let $\by$ be a vector in $\mathbb{Z}^n$. We observe that $(\by \modulo \Lambda) \modulo p$ (where the first modulo operation is performed with respect to $\cV$) is equal to $\by^* \modulo \cC$ (where the modulo operation is performed with respect to $\cV^*$). That is, the order of the modulo operations can be exchanged.
We conclude that for a lattice obtained by Construction~A, defining $\cV$ is equivalent to defining $\cV^*$. We may therefore focus on the
latter task.

We introduce now the technique by which we partition $\ZZ^n$ for a given lattice $\Lambda$ and
target PDF $f_W(w)$. We rely on typicality  (see, e.g., \cite{CoverBook}) and
use the following notation:
\begin{itemize}
\item $A_{\epsilon}^{(n)}(W)$: The set of  vectors in $\ZZ_p^n$ that are $\epsilon$-typical to $P_W$.
\item $A_{\epsilon,W}^{(n)}(X,Y)  \Ddef \{\bx,\by| (\by-\bx) \modulo p \in
A_{\epsilon}^{(n)}(W)\}$: The set of all pairs of vectors in $\ZZ_p^n$ such that their {\em difference} modulo
$p$ is $\epsilon$-typical to the PDF of $W$.
\end{itemize}

We go over all points $\by \in \mathbb{Z}_p^n$. For any $\by$ not yet associated to a cell we
associate it according to the following two possibilities:
\begin{enumerate}
\item 
There is no ${\bf X_i} \in \cC$ such that $({\bf X_i},\by) \in
A_{\epsilon,W}^{(n)}$: We arbitrarily associate $\by$ to  ${\bf X_0}=0$ and $\by \in \cV^*$.
\item 
There exists at least one codeword ${\bf X_i}$ such that $({\bf X_i},\by) \in A_{\epsilon,W}^{(n)}$: We choose one such codeword and add $(\by-{\bf X_i}) \modulo p$ to $\cV^*$.
\end{enumerate}
For any such vector $\by$, we also associate all the ``coset members" $\by-{\bf X_i} + \bX_j \modulo p$
to the their respective cells $\cV^*+ \bX_j \modulo p$.
Thus, in each step we first associate the vector $\br = (\by-{\bf X_i}) \modulo p$ to the basic cell and then map the $p^k-1$ coset members.
We then apply the procedure again until all vectors $\by \in \ZZ_p^n$
have been associated.
This procedure is depicted in Figure~\ref{code_mapping}.

\section{Proof of Theorem~\ref{DiscreteShaping}}
\label{sec:proof} We now show
that indeed we obtain an ensemble of lattices and associated
partitions such that the LQN has the desired properties, for almost all members of the ensemble.
The main steps are as follows.
Lemma~\ref{xi}
shows that for any vector in $\ZZ_n^p$ the probability to find a 
codeword such that their difference (modulo $p$) is typical to $W$ goes to one as the
dimension goes to infinity, with a proper choice for the rate of
the codebooks. We conclude that there exists a specific
sequence of codebooks that can match almost every point of $\ZZ_p^n$
and then show that such a sequence yields the desired LQN.
In the sequel we form the partitioning letting $\epsilon$ decrease with dimension $n$ as:
\begin{eqnarray}
\epsilon = \frac{1}{n}. \label{EDef} 
\end{eqnarray}
Any other vanishing function of $n$ would be appropriate.

\subsection{Almost all points are matchable}

For every code $\cC$, let $\cC'=\cC+\bD \modulo p$ be a {\em randomly shifted} version of the
codebook (i.e., $\cC'$ is a random coset code)  where $\bD$ is a random vector uniformly distributed over $\ZZ_p^n$.
We denote the codewords of $\cC'$ by $\bX'_i$, $i=1,\ldots,p^k$. Thus, $\bX_i'=\bX_i + \bD \modulo p$.

For a given $\by \in \mathbb{Z}_p^n$, we call the event in which there is no codeword in $\cC'$ such that its
difference from $\by$ modulo $p$ is typical to $W$, as a ``bad event". Defining the indicator random variable,
\begin{eqnarray*}
\zeta(\by) = \left\{
\begin{array}
{ll} 0, &  \nexists \bX'_i \in \cC' \ \ s.t \ \ (\bX'_i,\by) \in A_{\epsilon,W}^{(n)}(X,Y) \\ 1, & o.w
\end{array}
\right. ,
\end{eqnarray*}
a bad event amounts to the event $\zeta(\by)=0$. The next lemma shows that with a proper choice of code rate, such ``bad events" are rare.
\begin{lemma}
\label{xi} Let $\cC$ be a linear codebook of rate $R$ drawn from the random ensemble defined
above and let $\cC'$ be the induced random coset code. Let $\by \in \mathbb{Z}_p^n$ be any given vector. 
Then, for a rate satisfying
\begin{equation}
R \geq \log p-H(W)+2\epsilon \label{rateeq}
\end{equation}
we have
\begin{eqnarray}
\lim_{n \rightarrow \infty} \Pr(\zeta(\by)=0) = 0,
\end{eqnarray}
where the probability is averaged over all codebooks and over all
shifting random vectors, and $\epsilon$ as defined in (\ref{EDef}).
\end{lemma}
\noindent The proof is given in Appendix~\ref{sec:proof_lemma}.

Denote by $N_{\bY}$ the number of vectors $\by \in \ZZ_p^n$ that can be matched. We note that
\begin{eqnarray*}
E[N_{\bY}] =
\Pr(\zeta(\by)=1),
\end{eqnarray*}
where the expectation is over all codebooks and all shifting random vectors. By Lemma~\ref{xi}, taking $n$ to infinity we get
\begin{eqnarray}
\lim_{n \rightarrow \infty}\frac{E[N_{\bY}]}{|\ZZ_p^n|} &=& \lim_{n \rightarrow \infty}
\Pr(\zeta(\by)=1) \nonumber = 1.
\end{eqnarray}
Note that this result applies also to the original (non-shifted) ensemble (and for
any other constant-shifted ensemble) due to symmetry. Thus, $E[N_{\bY}]=E[N_{\bY} | \bD=\bd]$ for {\em any} shift vector $\bd$, where the expectation on the r.h.s is only over the lattice ensemble.

An immediate consequence is that there
exists a specific sequence of codebooks $\cC_n$ for which
\begin{eqnarray}
\lim_{n \rightarrow \infty}\frac{N_{\bY}}{|\ZZ_p^n|} = 1.
\label{ratio0}
\end{eqnarray}
We focus our attention on such a sequence and consider the corresponding sequence of fundamental
regions. 
Let us denote the set of matchable sequences in $\cV^*$ by $\cV^*_g$ and the non-matchable by
$\cV^*_b$. From the symmetrical construction of the cells and from (\ref{ratio0}) it follows that:
\begin{eqnarray}
\label{ratio} \lim_{n \rightarrow \infty}  \frac {|\cV^*_b|} {|\cV^*|} = 0.
\end{eqnarray}
Thus almost all points in $\cV^*$ are typical to $W$.

\subsection{Convergence in divergence}
\label{subsec:pdfconv} We now show that the resulting LQN reduced modulo $p$, $\bU^*$,
asymptotically approaches the desired distribution. The construction
suggested above
creates $p^{k}$ cells, each with $p^{n-k}$ elements. 
Thus, $\bU^*$ assumes one of $p^{n-k}$ values with equal probability. We now relate the entropy of
$W$ to the volume of a cell.
We take the rate of the code to satisfy (\ref{rateeq}) with equality, i.e.,
\begin{eqnarray}
R = \log p-H(W) + 2\epsilon,
\end{eqnarray}
where $\epsilon$ is defined in (\ref{EDef}). We thus have
\begin{eqnarray*}
\frac{1}{n} \log  p^k = \log  p - H(W) + 2\epsilon,
\end{eqnarray*}
or equivalently
\begin{eqnarray*}
2^{n(H(W)-2\epsilon)} = p^{n-k}.
\end{eqnarray*}
We observe that
\begin{itemize}
\item For each $\by \in \cV^*$,
\begin{eqnarray}
\label{UPr}
P_{\bU^*}(\bU^* = \by) = 2^{-n(H(W)-2\epsilon)}.
\end{eqnarray}
\item For $\by \in \cV^*_g$,
\begin{eqnarray}
\label{WgPr}
P_{\bW}(\bW = \by) \geq 2^{-n(H(W)+\epsilon)},
\end{eqnarray}
by the definition of (weak) typicality (i.e.Definition (3.6) in \cite{CoverBook}).
\item For $\by \in \cV^*_b$, $\Pr(\bW = \by) \geq (a_{\rm min})^n$. Defining
\begin{eqnarray}
\alpha \Ddef -\log_{2}a_{\rm min} - H(W) \label{alpha},
\end{eqnarray}
it follows that
\begin{eqnarray}
\label{WbPr}
\Pr(\bW = \by) \geq 2^{-n(H(W)+\alpha)}
\end{eqnarray}
for any $\by \in \cV^*_b$.
\end{itemize}
We thus have,
\begin{eqnarray*}
  &&D(\bU^*||\bW) \\
  & = & \sum_{\by \in \cV^*} P_{\bU^*}(\bU^*=\by) \log \left(\frac{P_{\bU^*}(\bU^*=\by)}{P_{\bW}(\bW=\by)}\right) \\
  & \leq & \sum_{\by \in \cV^*_g} 2^{-n(H(W)-2\epsilon)} \log \left(\frac{2^{-n(H(W)-2\epsilon)}}{2^{-n(H(W)+\epsilon)}}\right) \\
  & + & \sum_{\by \in \cV^*_b} 2^{-n(H(W)-2\epsilon)} \log \left(\frac{2^{-n(H(W)-2\epsilon)}}{2^{-n(H(W)+\alpha)}}\right)\\
  & = & \left(2^{n(H(W)-2\epsilon)}-|\cV^*_b|\right) \cdot 2^{-n(H(W)-2\epsilon)} \cdot n(2\epsilon+\epsilon)\\
  & + & |\cV^*_b| \cdot 2^{-n(H(W)-2\epsilon)} \cdot n(\alpha+2\epsilon) \\
  & = & |\cV^*_b| \cdot 2^{-n(H(W)-2\epsilon)} \cdot n(\alpha-\epsilon) + 3 \epsilon n.
\end{eqnarray*}
Dividing both sides by $n$, we get
\begin{eqnarray}
\frac{1}{n}D(\bU^*||\bW) & \leq & 3\epsilon + \frac {|\cV^*_b|} {|\cV^*|} \cdot (\alpha-\epsilon) \\
& = & \epsilon^{*}, \label{EStar}
\end{eqnarray}
where
\begin{eqnarray}
\epsilon^{*} & = & 3\epsilon + \frac {|\cV^*_b|} {|\cV^*|} \cdot (\alpha-\epsilon).
\label{eps4}
\end{eqnarray}
From (\ref{ratio}) it follows that the second term of the r.h.s. of (\ref{eps4}) vanishes as $n$ goes to infinity. In
addition, it is clear (by its definition) that $ \epsilon$ vanishes  as well as  $n$
goes to infinity and hence the same applies to $\frac{1}{n}D(\bU^*||\bW)$. This completes the proof of Theorem~\ref{DiscreteShaping}.


\section{Tying it all together}
\label{sec:tying}
We turn to proving Theorem~\ref{ContinuousShaping}. We show how we may generate any desired continuous distribution, subject to mild regularity conditions, by building on the
results derived for the discrete case. Let us denote the desired permissible continuous LQN by $W^c$ and its PDF by $f_{W^c}(\cdot)$ .
First, divide $\cA$ into small enough $\Delta$-size intervals, where $\Delta$ will be a constant value such that
\begin{eqnarray}
\frac{|\cA|}{\Delta}=\frac{2A}{\Delta}=p
\end{eqnarray}
is a prime number. The larger the value of $p$, the more refined the approximation for $f_{W^c}(\cdot)$ will be.
We then use the following steps to construct the lattice and lattice partition:
\begin{itemize}
\item Define the folded random variable $W^{c,*}$ by,
\begin{eqnarray}
W^{c,*} = W^c \modulo [0,2A].
\end{eqnarray}
\item Define the quantized random variable $W$ by,
\begin{eqnarray}
W=y \ \  \mbox{if} \ \ y \Delta \leq W^{c,*} < (y+1)\Delta,
\label{eq:W_def}
\end{eqnarray}
where $y$ takes values in $\mathbb{Z}_p=\{0,1,\ldots,p-1\}$.
Denote the PDF of $W$ by $f_W(\cdot)$.
\item Generate the sequence of lattices and lattice partitions, $\cV$, as described in the previous section such that the associated LQN approaches an i.i.d. distribution with marginal PDF $f_W(w)$.
\item Scale the lattice by $\Delta$, i.e
\begin{eqnarray}
    \Lambda^c = \Delta \cdot \Lambda.
\end{eqnarray}
\item Define the continuous fundamental region, $\cV^c$, by,
\begin{eqnarray}
\cV^c = \Delta \cdot \cV + \cI_{\Delta}^n.
\end{eqnarray}
where
\begin{eqnarray}
\cI_{\Delta}^n=[0,\Delta)^n.
\end{eqnarray}
That is, we scale $\cV$ by a factor of $\Delta$ and add a $\Delta$-size cube to each element.
\end{itemize}

\begin{figure}[htb]
\epsfysize=1.1in
\includegraphics[width=0.9\columnwidth]{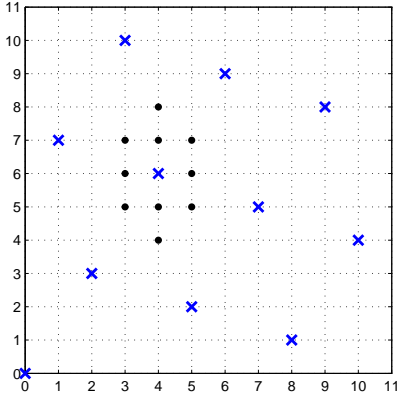}
\caption{Discrete lattice and lattice partition with $p=11$}
\label{discrete_partition}
\end{figure}

\begin{figure}[htb]
\epsfysize=1.1in
\includegraphics[width=0.9\columnwidth]{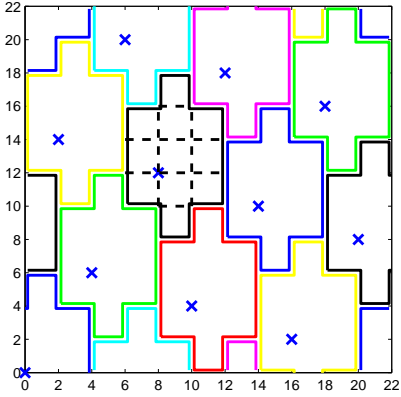}
\caption{Equivalent continuous lattice and lattice partition with $\Delta=2$}
\label{continuous_partition}
\end{figure}

These steps are exemplified in Figures~\ref{discrete_partition} and \ref{continuous_partition}. Figure~\ref{discrete_partition} depicts a discrete lattice with $p=11$ where the codewords are designated with `x`. The cell of some codeword is designated with dots. Figure~\ref{continuous_partition} depicts the equivalent continuous lattice with $\Delta=2$. Here the codewords are designated with `x` while the cells are drawn with a solid line.

Let $\bU^c \sim {\rm Unif}(\cV^c)$ be the resulting LQN.
In Appendix~\ref{sec:proof_of_cont} we show that indeed the construction yields an LQN $\bU^c$  which PDF that can approximate $f_{W^c}(\cdot)$ to any desired degree in a Kulback-Leibler divergence sense, thus completing the proof of Theorem~\ref{ContinuousShaping}.

\section{Simulation Results}
\label{sec:sim} In this section we demonstrate the theoretical results
via simulation. Since for complexity reasons we are limited to using only small dimensions, we
replaced the typicality criterion with the maximum likelihood
criterion. In addition, instead of choosing a lattice at random, we generated at random $100$ codebooks and picked the codebook that maximizes $D(\bU||\bW)$.

\begin{figure}[htb]
\includegraphics[width=.9\columnwidth]{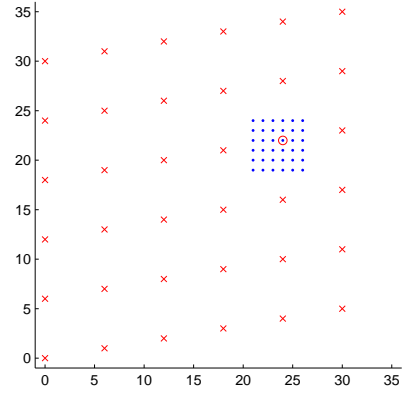}
\caption{Lattice partition for $W_1$} \label{step_6_lattice}
\end{figure}

\begin{figure}[htb]
\includegraphics[width=.9\columnwidth]{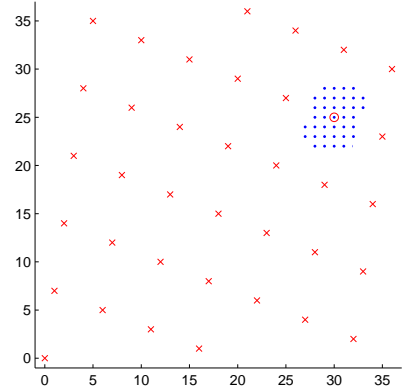}
\caption{Lattice partition for $W_2$} \label{trapeze_lattice}
\end{figure}

\begin{figure}[htb]
\epsfysize=1.1in
\includegraphics[width=.9\columnwidth]{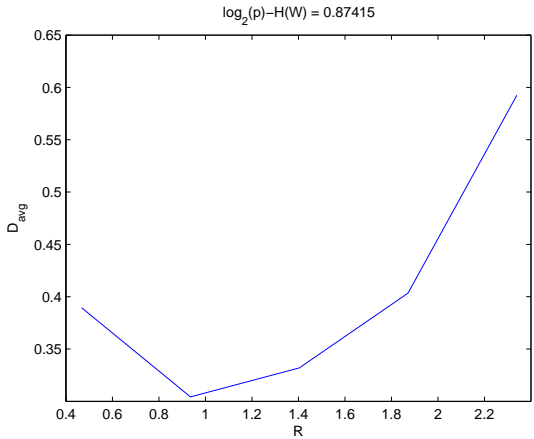}
\caption{Relative entropy for different values of $R$, for $W_3$.} \label{RvsD}
\end{figure}

\begin{figure}[htb]
\includegraphics[width=.85\columnwidth]{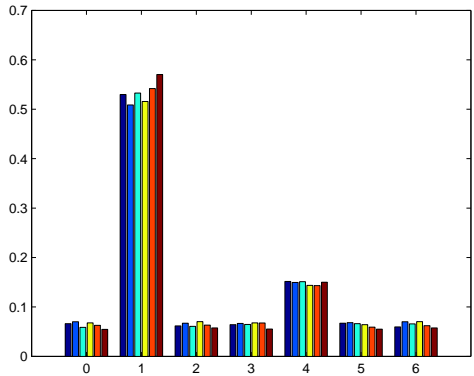}
\caption{PDFs of the obtained marginal distributions, for $W_3$.} \label{pdfs}
\end{figure}

\begin{figure}[htb]
\epsfysize=1.1in
\includegraphics[width=.9\columnwidth]{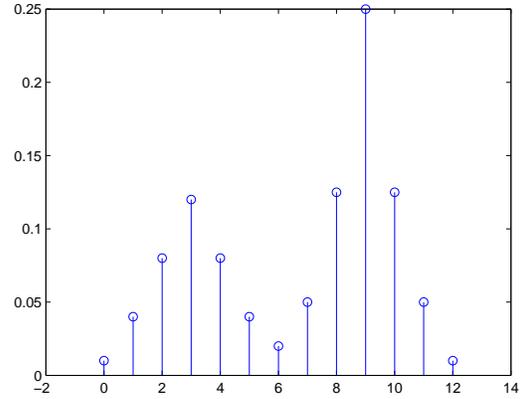}
\caption{Desired PDF $f_{W_4}$.} \label{WTriangle}
\end{figure}

\begin{figure}[htb]
\epsfysize=1.1in
\includegraphics[width=.9\columnwidth]{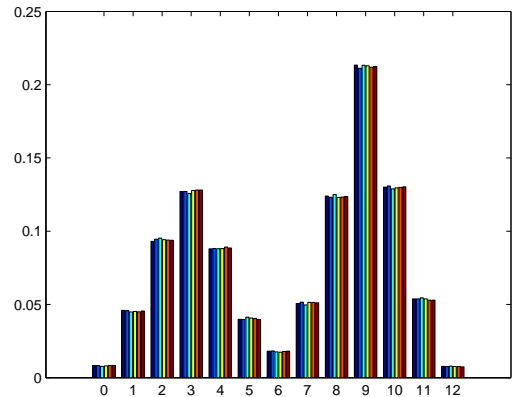}
\caption{PDFs of the obtained marginal distributions, for target distribution $f_{W_4}$.} \label{UTriangle}
\end{figure}

We considered the following cases:
\begin{itemize}
\item $p=37, n=2, k=1$,
\[
f_{W_1}(w)=\left\{
\begin{array}
{ll}
0.999/6, & w \in \{0,1,2, 34,35,36\} \\
0.001/31, & o.w \\
\end{array}
\right.
\]
\item $p=37, n=2, k=1$,
\[
f_{W_2}(w)=\left\{
\begin{array}
{ll}
0.1427, & w \in \{0,1,2, 35,36\} \\
0.0951, & w \in \{3, 34\} \\
0.0476, & w \in \{4, 33\} \\
0.001/28, & o.w \\
\end{array}
\right.
\]
\item $p=7, n=6, k=1,...,5$,
\[
f_{W_3}(w)=\left\{
\begin{array}
{ll}
0.6, & w = 1 \\
0.15, & w = 4 \\
0.05, & w \in \{0, 2, 3, 5, 6\} \\
\end{array}
\right.
\]
\item $p=13, n=6, k=1$, $f_{W_4}(w)$ as depicted in Figure~\ref{WTriangle}.
\end{itemize}
Figure~\ref{step_6_lattice} shows the lattice partition that was obtained for the ``step like"
distribution $f_{W_1}(w)$. The codewords are designated with `x`. The cell of some codeword is designated with dots. As
expected, the lattice cell has a shape of a square and the divergence from the desired distribution is $0$. 
Figure~\ref{trapeze_lattice} shows the lattice partition obtained for $f_{W_2}(w)$.
Figures~\ref{RvsD} and \ref{pdfs} refer to the distribution of the third case. Figure~\ref{RvsD}
depicts the relative entropy corresponding to each value of the rate (corresponding to $k=1,\ldots,5$). The optimal value was obtained
for the rate that was the closest to $\log p - H(W)$ as should be expected.
Figure~\ref{pdfs} depicts the marginal distribution of each element of the vector $\bU$ that was
obtained using the optimal rate,  which is in good agreement with $f_{W_3}(w)$.
Finally, Figure~\ref{WTriangle} depicts the desired PDF $f_{W_4}$.
The simulation was run for dimension $n=6$ and with $k=1$ (for $n=6$ this is the optimal value for $k$). 
Figure~\ref{UTriangle} depicts the marginal distribution of the obtained LQN which is in good agreement with the desired one. 

\section{Summary}
It was demonstrated that subject to mild regularity conditions, lattice quantization noise may approach (asymptotically in the dimension)  
quite general distributions, with a proper choice of lattice and partitioning.

\section{APPENDIX}  

\setcounter{section}{0}  
\renewcommand{\thesection}{\Alph{section}}

\section{Proof Of Lemma~\ref{xi}}
\label{sec:proof_lemma}
We note that by standard arguments the codewords of $\cC'$ are pairwise independent and uniformly
distributed over $\ZZ_p^n$.
Define the indicator random variables
\begin{eqnarray*}
\gamma_i(\by) = \left\{
\begin{array}
{ll} 1, & ({\bf X_i}', \by) \in A_{\epsilon,W}^{(n)}(X,Y) \\ 0, & o.w
\end{array}
\right.
\end{eqnarray*}
and note that they are also pairwise independent. We have
\begin{eqnarray*}
E\left[\gamma_i(\by) \right] = b, & \sigma^2\left[\gamma_i(\by)\right] = b(1-b)
\end{eqnarray*}
and
\begin{eqnarray*}
E[\gamma_i(\by) \gamma_j(\by)] = \left\{
\begin{array}
{ll} \sigma^2+b^2, & i=j \\
b^2, & i \neq j
\end{array}
\right.
\end{eqnarray*}
where
\begin{eqnarray*}
b = \Pr\left(({\bf X_i}',\by) \in A_{\epsilon,W}^{(n)}(X,Y)\right)
\end{eqnarray*}
Note that $b$ is independent of the vector $\by$. Since $\by$ plays no role in the
analysis, we thus omit it from the notation and use $\gamma_i=\gamma_i(\by)$ below.

We note that $\bX_i'$ is drawn uniformly over $\ZZ_p^n$. Therefore,
the difference $\bE=\bX_i'-\by \modulo p$ is also distributed uniformly over $\ZZ^n_p$.
Thus,
\begin{eqnarray*}
b & = &  \frac {|A_{\epsilon}^{(n)}(W)|} {|\ZZ^n_p|} \\
  & \geq & \frac {(1-\epsilon) 2^{n(H(W)-\epsilon)}} {p^n} \\
  & = & (1-\epsilon) 2^{-n(\log p-H(W)+\epsilon)},
\end{eqnarray*}
where the inequality follows from the definition of typicality (i.e.,  by Theorem~3.1.2 in \cite{CoverBook}).

We denote by $M=p^k$ the total number of codewords in $\cC$
and bound the probability of a ``bad event" (no match) by
\begin{eqnarray}
\Pr(\zeta(\by)=0) & = & \Pr\left(\sum_{i=1}^{M} \gamma_i = 0\right)  \nonumber \\
   & = & \Pr\left(\frac{1}{M} \sum_{i=1}^{M} \gamma_i - b = -b \right) \nonumber\\
   & \leq & \Pr\left(\left|\frac{1}{M} \sum_{i=1}^{M} \gamma_i - b \right| \geq b \right) \nonumber\\
   & \leq & \frac{1}{b^2} E\left[\left(\frac{1}{M} \sum_{i=1}^{M} \gamma_i - b\right)^2\right] \nonumber\\
   & = & \frac{1}{b^2} \left(\frac{1}{M^2} \sum_{i=1}^{M}\sum_{j=1}^{M}E[\gamma_i \gamma_j]-b^2\right) \nonumber\\
   & = & \frac{1}{b^2 M^2} \sum_{i=1}^{M}\sum_{j=1}^{M}\left(E[\gamma_i \gamma_j]-b^2\right) \nonumber\\
   & = & \frac{1}{b^2 M^2} \left[M\left(E[\gamma_1^2]-b^2\right) \right. \nonumber\\
   & + & \left. (M^2-M)\left(E[\gamma_1\gamma_2]-b^2\right)\right] \nonumber\\
   & = & \frac{M \sigma^2} {b^2M^2} = \frac{b(1-b)}{b^2M} = \frac{1-b}{bM} \nonumber\\
   & \leq & \frac{1}{bM} \leq \frac
   {1}{(1-\epsilon)2^{nR}2^{\left(-n\log p-H(W)+\epsilon \right)}}\nonumber\\
   & = & (1-\epsilon)^{-1}2^{-n\left[R-\left(\log p-H(W)+\epsilon \right)\right]}
\end{eqnarray}
where the fourth transition is due to Chebyshev's inequality using the fact that
$E\left[\frac{1}{M} \sum_{i=1}^{M} \gamma_i\right]=b$. Thus, for any $\by \in \ZZ_p^n$, $\Pr(\zeta(\by)=0) \rightarrow 0$ as $n$ goes to infinity.

\section{Proof Of Continuous Case}
\label{sec:proof_of_cont} In this Appendix we show that the construction of lattices and lattice partitions as has been described in Section \ref{sec:tying} allows us to approach the desired distribution as closed as desired.

We note that each $\bt \in \cV^{c,*}$ can be
uniquely written as $\bt=\Delta\by+\bl$ where $\by \in \cV^*$ and $\bl
\in \cI_{\Delta}^n$. For every $\bt$ we denote the unique $\by$ associated with it by $\byt$.

Consider the random vector $\bW$ defined in (\ref{eq:W_def}). That is, each component $W_i$ is
the quantization of $W^{c,*}_i$. We further note that
\begin{eqnarray}
f_{\bW^{c,*}}(\bt) &=& \Pr(\bW^{c,*} \in \Delta\byt+\cI_{\Delta}^n) \nonumber\\
& \times & f_{\bW^{c,*}}(\bt | \bW^{c,*} \in \Delta\byt+\cI_{\Delta}^n) \nonumber\\
& = & \Pr(\bW=\byt) \cdot f_{\bW^{c,*}}(\bt|\bW=\byt).
\end{eqnarray}

Let $\cV^{c,*} = \cV^c \modulo [0,2A]$ be the ``folded" continuous fundamental region and
let $\bU^{c,*} \sim {\rm Unif}(\cV^{c,*})$ be the ``folded" LQN. Note that $\bU^{c,*} = \bU^c \modulo [0,2A]$.
Define the quantized random variable $U$ by,
\begin{eqnarray}
U=y \ \  \mbox{if} \ \ y \Delta \leq U^{c,*} < (y+1)\Delta,
\label{eq:U_def}
\end{eqnarray}
where $y$ takes values in $\mathbb{Z}_p=\{0,1,\ldots,p-1\}$ and
consider the random vector $\bU$ where each component $U_i$ is the quantization of $U^{c,*}_i$.
Note that
\begin{eqnarray}
f_{\bU^{c,*}}(\bt) = \Pr(\bU=\byt) \cdot f_{\bU^{c,*}}(\bt|\bU=\byt).
\end{eqnarray}

Define $\eta$ by:
\begin{eqnarray}
\label{etadef}
\eta^{-1} &=& \arg\!\min_{t,y} f_{W^{c,*}}(t|W=y) \\
&=& \arg\!\min_{t,y} \frac{f_{W^{c,*}}(t)}{\int_{y}^{y+\Delta} f_{W^{c,*}}(t)dt} \nonumber \\
&\geq& \arg\!\min_{y} \frac{1}{\Delta} \cdot \frac{\arg\!\min_{t \in \cJ(t,\Delta)}{f_{W^{c,*}}(t)}}{\arg\!\max_{t \in \cJ(t,\Delta)}{f_{W^{c,*}}(t)}} \nonumber \\
&=& \frac{1}{\Delta} \cdot r
\end{eqnarray}
where
\begin{eqnarray}
y \in \{0,...,p-1\}
\end{eqnarray}
and where
\begin{eqnarray}
\cJ(t,\Delta) = \left[y \Delta, (y+1)\Delta\right)
\end{eqnarray}
and
\begin{eqnarray}
r = \arg\!\min_{y} \frac{\arg\!\min_{t \in \cJ(t,\Delta)}{f_{W^{c,*}}(t)}}{\arg\!\max_{t \in \cJ(t,\Delta)}{f_{W^{c,*}}(t)}} \nonumber \\
\end{eqnarray}
We observe that
\begin{itemize}
\item For each $\bt \in \cV^{c,*}$,
\begin{eqnarray*}
f_{\bU^{c,*}}(\bt) &=& \Pr(\bU=\byt) \cdot f_{\bU^{c,*}}(\bt|\bU=\byt) \\
                   &=& 2^{-n(H(W)-2\epsilon)}\cdot \Delta^{-n}\\
                   &=& 2^{-n(H(W)-2\epsilon+\log \Delta)}
\end{eqnarray*}
where $\epsilon$ is defined in (\ref{EDef}) and where we used (\ref{UPr}) and the fact that $\bt$ given $\byt$ is uniformly distributed over $\cI_{\Delta}^n$.
\item For each $\bt \in \cV^{c,*}$ such that $\byt \in \cV^*_g$,
\begin{eqnarray*}
f_{\bW^{c,*}}(\bt) &=& \Pr(\bW=\byt) \cdot f_{\bW^{c,*}}(\bt|\bW=\byt) \\
                   & \geq & 2^{-n(H(W)+\epsilon)}\cdot \eta^{-n}\\
                   &=& 2^{-n(H(W)+\epsilon+\log \eta)}
\end{eqnarray*}
where $\epsilon$ is defined in (\ref{EDef}) and where we used (\ref{WgPr}) and (\ref{etadef}) to get the inequality.
\item For each $\bt \in \cV^{c,*}$ such that $\byt \in \cV^*_b$,
\begin{eqnarray*}
f_{\bW^{c,*}}(\bt) &=& \Pr(\bW=\byt) \cdot f_{\bW^{c,*}}(\bt|\bW=\byt) \\
                   &\geq& 2^{-n(H(W)+\alpha)}\cdot \eta^{-n}\\
                   &=& 2^{-n(H(W)+\alpha+\log \eta)}
\end{eqnarray*}
where $\alpha$ is defined in (\ref{alpha}) and where we used (\ref{WbPr}) and (\ref{etadef}). 
\end{itemize}

Using the sequence of lattices proposed in Theorem~\ref{DiscreteShaping} we obtain,
\begin{eqnarray}
&& D(\bU^c||\bW^c) \nonumber\\
  &=& \int_{\cV^c}f_{\bU^c}(\bt)\log\frac{f_{\bU^c}(\bt)}{f_{\bW^c}(\bt)} d{\bt} \nonumber\\
  &=& \int_{\cV^{c,*}}f_{\bU^{c,*}}(\bt)\log\frac{f_{\bU^{c,*}}(\bt)}{f_{\bW^{c,*}}(\bt)} d{\bt} \nonumber\\
  &=& \sum_{\by \in \cV^*} \int_{\bl \in \cI_{\Delta}^n} f_{\bU^{c,*}}(\Delta\by+\bl) \log
  \frac{f_{\bU^{c,*}}(\Delta\by+\bl)}{f_{\bW^{c,*}}(\Delta\by+\bl)} d\bl \nonumber\\
  &\leq& \sum_{\by \in \cV^*_g} \int_{\bl \in \cI_{\Delta}^n} 2^{-n(H(W)-2\epsilon+\log \Delta)} \nonumber\\
  & \times & \log \frac{2^{-n(H(W)-2\epsilon+\log \Delta)}}{2^{-n(H(W)+\epsilon+\log \eta)}} d\bl\nonumber\\
  &+& \sum_{\by \in \cV^*_b} \int_{\cI_{\Delta}^n} 2^{-n(H(W)-2\epsilon+\log \Delta)} \nonumber\\
  & \times &\log \frac{2^{-n(H(W)-2\epsilon+\log \Delta)}}{2^{-n(H(W)+\alpha+\log \eta)}} d\bl \nonumber\\
  &=& (2^{n(H(W)-2\epsilon)}-|\cV^*_b|)2^{-n(H(W)-2\epsilon)}\nonumber\\
  &\times& n(2\epsilon-\log \Delta+\epsilon+\log \eta)\nonumber\\
  &+& |\cV^*_b| \cdot 2^{-n(H(W)-2\epsilon)} \cdot
  n(2\epsilon-\log \Delta+\alpha+\log \eta)\nonumber\\
  &=& n(3\epsilon+\log \left(\frac{\eta}{\Delta}\right)) \nonumber\\
  &+& |\cV^*_b| \cdot 2^{-n(H(W)-2\epsilon)}\cdot n(\alpha-\epsilon)
\end{eqnarray}
Dividing both sides by $n$, we get
\begin{eqnarray}
\frac{1}{n}D(\bU^c||\bW^c) & \leq & \epsilon^{*} +
\log \left(\frac{\eta}{\Delta}\right),
\end{eqnarray}
where $\epsilon^{*}$ is defined in (\ref{EStar}). Therefore,
\begin{eqnarray}
\lim_{n \rightarrow \infty} \frac{1}{n}D(\bU^c||\bW^c) & \leq & \log \left(\frac{\eta}{\Delta}\right) \nonumber \\
& \leq & \log \left(\frac{\Delta}{\Delta r}\right) = -\log (r)
\label{eq:lasteq}
\end{eqnarray}
It remains to choose $\Delta$ such that $r$ is close enough to $1$.
Note that since $f_{W^c}$ is permissible, it is continuous and limited to the closed interval $\cA$. Therefore, $f_{W^c}$ is uniformly continuous, i.e, for any $\theta > 0$ exists $\Delta > 0$ such that for any $x_1,x_2 \in \cA$ satisfying $|x_1-x_2| \leq \Delta$, it follows that $|f_{W^c}(x_1)-f_{W^c}(x_2)| \leq \theta$.
Therefore, $r$ can be lower bounded by
\begin{eqnarray}
r \geq \frac{a_{\rm min}} {a_{\rm min} + \theta}
\end{eqnarray}
where $a_{\rm min}$ is defined in (\ref{FuncCond}).
If we choose
\begin{eqnarray}
\theta = a_{\rm min}(2^\xi-1)
\end{eqnarray}
and 
set $\Delta=\theta$, then $r \geq 2^{-\xi}$ and by (\ref{eq:lasteq}),
\begin{eqnarray}
\limsup_{n \rightarrow \infty} \frac{1}{n}D(\bU^c||\bW^c) = \xi.
\end{eqnarray}

\section{Proof Of Corollary}
\label{sec:proof_of_corollary}
We prove Corollary \ref{OneDimensionDisc} for the discrete case. \\
\begin{proof} We first
use the chain rule for relative entropy (see Theorem 2.5.3 in
\cite{CoverBook}):
\begin{eqnarray}
D(\bU||\bW) &=& D(P_{\bU}(\bx)||P_{\bW}(\bx)) \nonumber \\
            &=& \sum_{i=1}^{n}D(P_{\bU}(x_i|x_1^{i-1})||P_{\bW}(x_i|x_1^{i-1}))
\end{eqnarray}
Using Theorem 2.7.2 in \cite{CoverBook} we get
\begin{eqnarray}
&&D(P_{\bU}(x_i|x_1^{i-1})||P_{\bW}(x_i|x_1^{i-1})) \nonumber \\
&=& \sum_{d_1^{i-1}} P_{\bU}(d_1^{i-1})\cdot \nonumber \\
&& D(P_{\bU}(x_i|x_1^{i-1}=d_1^{i-1})||P_{\bW}(x_i|x_1^{i-1}=d_1^{i-1})) \nonumber \\
&\geq&
D\left(\sum_{d_1^{i-1}}P_{\bU}(d_1^{i-1})P_{\bU}(x_i|x_1^{i-1}=d_1^{i-1})||\right. \nonumber \\
&&\left.\sum_{d_1^{i-1}}P_{\bU}(d_1^{i-1})P_{\bW}(x_i|x_1^{i-1}=d_1^{i-1})\right) \nonumber \\
&=&D(P_{\bU}(x_i)||P_{\bW}(x_i)) \label{DinEq}
\end{eqnarray}
where the last transition is due to the fact that the elements of $\bW$
are i.i.d. Using (\ref{DinEq}) we get
\begin{eqnarray*}
D(\bU||\bW) \geq \sum_{i=1}^{n}D(P_{\bU}(x_i)||P_{\bW}(x_i)) =
\sum_{i=1}^{n}D(U_i||W_i)
\end{eqnarray*}
Finally, we conclude that using the same sequence of lattices
proposed in Theorem~\ref{DiscreteShaping} results in
\begin{eqnarray}
\lim_{n \rightarrow \infty} \frac{1}{n} \sum_{i=1}^{n}
D(U_i||W_i) \leq \lim_{n \rightarrow \infty} \frac{1}{n}
D(\bU||\bW)=0.
\end{eqnarray}
\end{proof}

\end{document}